\documentclass[aps,floatfix]{revtex4}
\usepackage{epsfig}
\usepackage{color}
\usepackage{amssymb}
\usepackage{amsmath}
\usepackage[ngerman,english]{babel}
\usepackage{multirow}
\usepackage{graphicx}
\usepackage{subfigure}

\newcommand{\etal}{{\it et al.}}
\newcommand{\pr}[4]{Phys. Rev. #1 {\bf #2}, #3 (#4)}
\newcommand{\hedp}[3]{High Energy Density Phys. {\bf #1}, #2 (#3)}
\newcommand{\astropj}[3]{Astrophys. J. {\bf #1}, #2 (#3)}

\begin{document}

\title{Pauli Blocking in Degenerate Plasmas and the Separable Potential Approach}

\author{Gerd R\"opke}
\affiliation{
Institut f\"ur Physik, Universit\"at Rostock, D-18051 Rostock, Germany}
\date{\today}

\begin{abstract}

The Mott effect describes the dissolution of bound states in a dense partially ionized plasma.
It happens when the ionization potential depression, owing to effects of correlation and degeneracy,
compensates the binding energy of the bound state. 
At high densities and moderate temperatures, the Pauli blocking becomes important and influences significantly the degree
of ionization in the region of degenerate plasmas. A quantum statistical approach is used where the total density is decomposed 
in an uncorrelated, "free" part and correlations, as a consequence of the cluster decomposition of the self-energy. 
The contribution of correlations to the total density is given by bound states and continuum correlations.
Exact solutions for a separable potential are compared to perturbation theory and numerical solutions of the in-medium 
Schr\"odinger equation. The in-medium scattering phase shifts are evaluated, and the role of continuum correlations is discussed.
The Pauli blocking of bound states and the density of states are considered for warm dense matter conditions.


\end{abstract}

\maketitle

\section{Introduction}

Warm dense matter (WDM) \cite{WarmDM} is an interesting region in the phase diagram of Coulomb systems, consisting of electrons and atoms in different states of ionization, at high densities (total ion densities $n_a \approx 10^{21} -10^{25}$ cm$^{-3}$) and not too high temperatures ($T \le 10^{3}$ eV) so that strong correlations and electron degeneracy are relevant. Matter under these conditions exists in astrophysical objects like planets or stars, but is 
also produced under laboratory conditions using pulsed power, 
high-power optical and free-electron-lasers or other methods of 
high-pressure experimental technique. To describe the properties of WDM,
concepts of plasma physics (such as the model of the partially ionized plasma (PIP)) and condensed matter physics
(such as Thomas-Fermi model or density-functional theory (DFT)) have to be 
extended beyond the limits of their conventional validity. 

In detail, the PIP model is valid in the low-density limit where the plasma is 
described as a mixture of electrons ($e$) and ions ($i$) with different degrees of 
ionization ($Z_i$) including neutral atoms ($Z_i=0$), which are found in 
different states of excitation. Charge neutrality is assumed so that for a single-component plasma
of the element $a$ with charge number $Z$ of the nucleus, $n^{\rm total}_e= Z n^{\rm total}_a$ holds. 
This gives for the free electron density $n_e$ the relation $n_e=\sum_i Z_i n_i$, 
where $n_i$ is the density of ions with charge number $Z_i$.
Correlation between these plasma constituents are 
considered as weak, the formation of higher aggregates such as molecules or clusters
may be included. An interesting quantity is the ionization degree $\bar Z = \sum_i Z_i n_i/n^{\rm total}_a$
which describes the average number of free electrons per atom.

The DFT model which is well-known in condensed matter physics 
is based on a mean-field concept where the electrons are treated quantum mechanically.
The ions are considered as 
classical particles which may be strongly correlated as calculated by molecular-dynamics (MD)
simulations. The electron correlations are treated in some approximation, using appropriate functionals 
for the energy density. There are recently intense work to improve both the PIP and DFT models to work out 
an approach to WDM which implements also these different limiting cases.

A quantum statistical approach \cite{KKER,KSK} is needed. As a fundamental relation, we have the
equation of state which relates the total densities of electrons $n_e^{\rm total}$ 
and nuclei $n_a^{\rm total}$ to the temperature $T=1/(k_B \beta)$ and the chemical 
potentials $\mu_e, \mu_a$,
\begin{equation}
\label{specn}
 n^{\rm total}_e(T,\mu_e,\mu_a)= \frac{1}{\Omega}\sum_1 \int_{-\infty}^\infty \frac{d \omega}{2 \pi}
\hat f_e(\omega) A_e(1,\omega)= \int_{-\infty}^\infty d \omega
\hat f_e(\omega) D_e(\omega)
%
\end{equation}
with the spectral function $A_e(1,\omega)$. The single-particle states are denoted by wave number vector and spin, 
$|1\rangle=|{\bf p}_1,\sigma_1 \rangle$, $\Omega$ is the system volume, and 
$\hat f_e(\omega)=[\exp (\beta \omega-\beta \mu_e)+1]^{-1}$
the Fermi distribution. 
A corresponding relation holds also 
for $n_a^{\rm total}$.
Both the variables $\mu_e, \mu_a$ are related to each other because of charge neutrality. 
From the spectral function another interesting quantity is obtained, the density of states (DOS) 
$D_e(\omega) =\Omega^{-1} \sum_1 A_e(1,\omega)/2 \pi$.

We use the method of Green functions \cite{KKER,KSK} 
which is based on perturbation theory, starting from free particles. 
Partial summation of Feynman diagrams is necessary
to implement the effects of correlations. 
In particular, we consider a separable potential approach \cite{EST} where 
the explicit solution of the in-medium Schr{\"o}dinger equation is available,
see \cite{SR87,R15} for the treatment of Pauli blocking in the electron-hole system and in nuclear matter.
As alternative, path-integral Monte-Carlo (PIMC)
simulations  \cite{Militzer,Dornheim} are applicable for strong correlations. 
However, it is restricted to the region of high temperatures and high densities because of the the fermion sign problem.

In this work, we follow Ref. \cite{PRE18} and analyze some concepts of the PIP which become problematic 
when describing high-density plasmas. In particular, we investigate the effects of degeneracy of the electrons.
A gas of free fermions is degenerate if $\Theta = T/T_{\rm Fermi}=(2 m_e k_BT/\hbar^2) \times (3 \pi^2 n_e)^{-2/3}\le1$.
Then, instead of the Boltzmann distribution, we have to use the Fermi distribution 
$f_e( p) =\hat f_e(E_p)$ with the free dispersion relation $E_p=\hbar^2 p^2/2m_e$. For the ideal fermion gas, the 
chemical potential $\mu_e$ is related to the density $n^{\rm id}_e=(1/\Omega) \sum_{\cal P} f_e( p)$ [free fermion spectral function 
$A^{\rm id}_e(1,\omega)= 2 \pi \delta(E_p-\omega)$], where with ${\cal P}=\{{\bf p},s\}$ the spin summation is  included.

Chemical equilibrium between the different components of the PIP is expressed as relations 
between the respective chemical potentials $\mu_i$, see Ref. \cite{PRE18}. In particular, the composition of the PIP
model is given by a coupled system of Saha equations, where in a plasma the free dispersion relation $E_p$ is replaced by the 
quasiparticle dispersion relation $E^{\rm quasi}(p)$ (including the mean-field energy shift). The ion densities are
\begin{equation}
\label{compni}
 n_i=\frac{1}{ \Lambda^{3}} \sigma_{i}(T)e^{\beta \mu_i}\,,\qquad \Lambda = [2 \pi \beta \hbar^2/M]^{1/2}
\end{equation}
(ion mass $M$) with the generalized Beth-Uhlenbeck formula for the intrinsic partition function, see \cite{PRE18}, 
\begin{equation}
\label{sigvir}
 \sigma_{i}(T)=\sum_\gamma\left[\sum^{\rm bound}_\nu \left(e^{\beta E_{i,\gamma, \nu}}-1\right)
+\frac{ \beta}{\pi} \int_0^\infty dE
e^{- \beta E}\left\{ \delta_{i, \gamma}(E)
 -\frac{1}{2} \sin [2  \delta_{i, \gamma}(E)]\right\} \right].
\end{equation}
The chemical potentials $\mu_i$ refer to the ground state of the ion with charge $Z_ie$. Different channels 
$\gamma$ (spin, angular momentum) for excitation of the ion are considered, 
the excitation energy $E_{i,\gamma, \nu}$ contains the intrinsic quantum number $\nu$ which covers 
the possible bound states as well as the scattering states, expressed by the scattering phase shift 
$\delta_{i, \gamma}(E)$.
In chemical equilibrium the relation $ \mu_i=\mu_{i+1}+\mu_e+I_i$ holds, where 
the ionization potential $I_i$ is the lowest 
excitation energy of the ionic ground state ($E_{i,0}$) to become ionized, i.e. $-I_i$ 
is the ground state energy of the ion $a_i$ relative to the continuum edge of $a_{i+1}+e$.

Compared to the isolated ion, in dense matter the energies $E_{i,\gamma, \nu}$ are modified 
(quasiparticle energies) because of mean-field shifts, correlations, and effects of degeneracy \cite{PRE18}.
In particular, the ionization potential $I^{(0)}_i$ of the isolated ion is reduced, and the difference 
$\Delta I_i=I_i^{(0)}-I_i$ is
denoted as ionization potential depression (IPD). Recent experiments with WDM 
\cite{Hoarty13,ciricosta12,ciricosta16,Vinko18,Fletcher14,Kraus16,CFT15} show an increasing interest 
in this quantity. Different many-particle effects contribute to the medium modification of the IPD, 
see \cite{Crowley14,Lin17,PRE18}. In this work, we analyze the Pauli blocking which becomes of interest in 
high-density WDM where the electrons are degenerate. The Pauli blocking reduces the binding energy 
and contributes to the dissolution of bound states (Mott effect), which is of relevance for the ionization degree $\bar Z$.

\begin{figure}[h]
  \centerline{\includegraphics[width=250pt,angle=0]{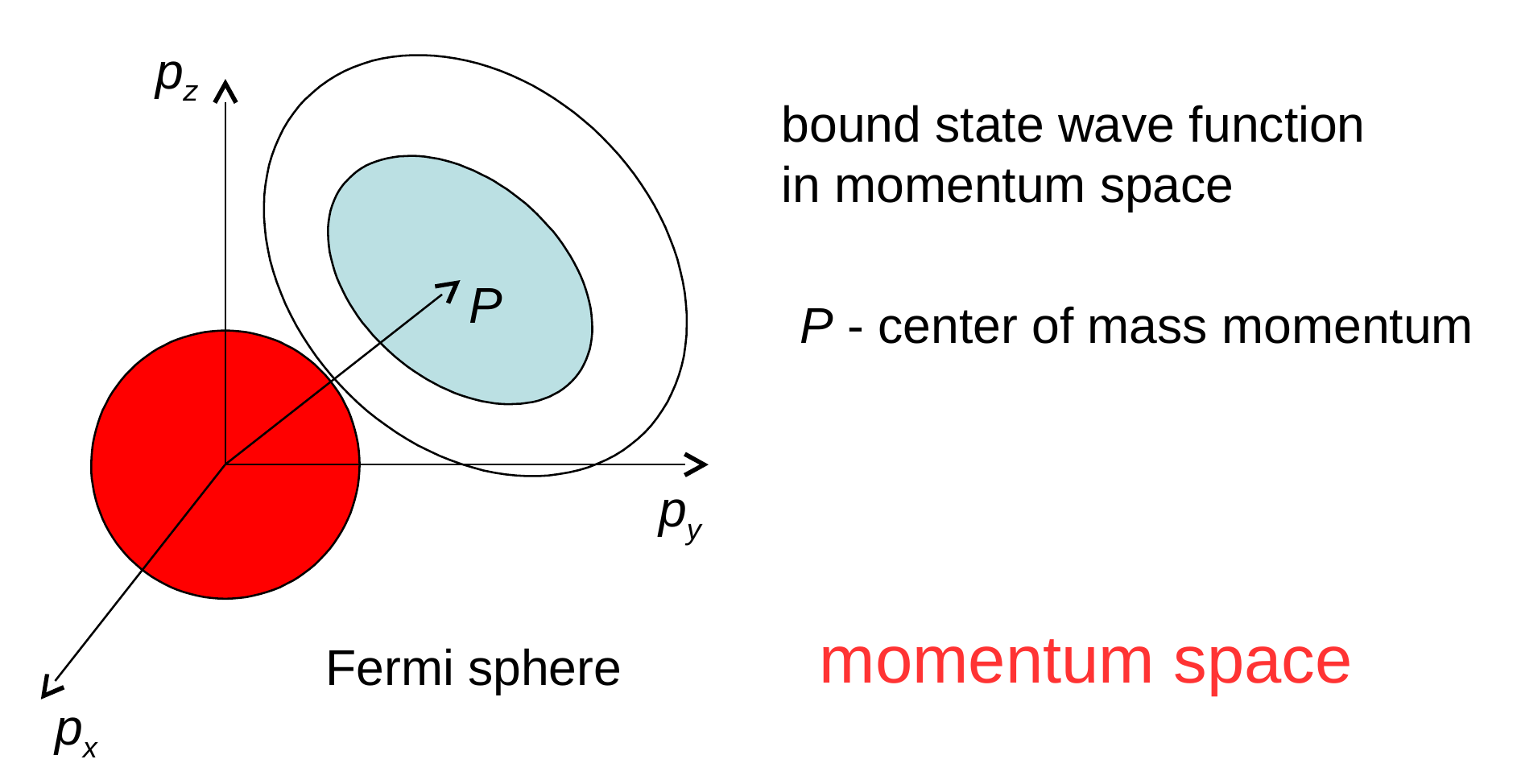}}
  \caption{Pauli blocking: Deformed bound-state wave function in phase space (momentum space). The Fermi sphere is already
occupied by the surrounding matter and cannot be used to form a bound state (Pauli principle), see Ref. \cite{Udo}.}
\label{fig:0}
\end{figure}

Whereas (Debye) screening is a well-known concept in plasma physics, Pauli blocking is less known. It is a consequence of the antisymmetrization of the fermionic wave function or the Pauli principle. 
A single-particle state cannot be occupied by more than one particle. A well-known example is the ground state of the ideal 
Fermi gas where all energy eigenstates below the Fermi energy are single-occupied. An additional particle can be added only in 
the phase space outside the Fermi sphere, see Fig. \ref{fig:0}. With respect to the formation of bound states, phase space 
must be available according to the bound-state wave function $\psi(\bf p)$ in momentum space. In a dense, degenerate system,
the momentum states within the Fermi sphere are already occupied and cannot be used to form the bound-state wave function
(Pauli blocking). As a consequence, the bound state energy cannot take its minimum value as in free space, but is shifted.
As directly seen from Fig. \ref{fig:0}, this shift is depending on the center-of mass momentum $\bf P$. 
The overlap of the free bound-state wave function with the Fermi sphere and consequently the deformation of the bound-state wave function in dense matter is a maximum for $P=0$.

\section{Correlations and Debye screening}

In a dense medium, the properties of the free single-particle states as well as of the bound states 
are influenced by the medium.
A systematic treatment of medium effects is obtained from Green-function theory \cite{KKER}, 
where the spectral function $A_e(1,\omega)$, Eq.  (\ref{specn}),  is 
related to the dynamical self-energy for which a cluster decomposition can be performed. 
With the self-energy, the quasiparticle concept can be introduced, as well-known in condensed matter theory.
As a result, the in-medium Schr{\"o}dinger
equation (or Bethe-Salpeter equation, BSE) for a few-particle system is obtained after partial summation of ladder diagrams. 
We consider the in-medium two-particle problem \cite{RKKKZ78} which is relevant for the excitation of the ion $a_i$ and the ionization $a_i \rightleftharpoons a_{i+1}+e$
\begin{eqnarray}
\label{BSE}
&&\left[E_e(p )+\Sigma_e(p,z)+E_{a_{i+1}}(k)+\Sigma_{a_{i+1}}(k,z)
\right] \psi^{i}_{n}({\bf p},{\bf k})\nonumber\\&&
+[1-f_e(p )\mp f_{a_{i+1}}(k)] \sum_{\bf q}V_{a_{i+1}, e}^{\rm eff}({\bf p},{\bf k},{\bf q},z) \psi^{i}_{n}({\bf p}+{\bf q},{\bf k}-{\bf q})=
E^{i}_{ n} \psi^{i}_{ n}({\bf p},{\bf k})
\end{eqnarray}
with the effective interaction
\begin{eqnarray}
\label{Veff}
&&V_{a_{i+1}, e}^{\rm eff}(1,2,{\bf q},z)= V_{a_{i+1}, e}(q)\left[1-\int_{-\infty}^\infty \frac{d \omega}{\pi}
{\rm Im}\, \varepsilon^{-1}(q,\omega+i0)\left[n_B(\omega)+1\right]\right.\nonumber \\&&
\left.\times\left[ \frac{1}{z-\omega-E_e(1)-E_{a_{i+1}}(2-{\bf q})}+\frac{1}{z-\omega-E_e(1+{\bf q})-E_{a_{i+1}}(2)}
\right]\right]\,.
\end{eqnarray}
$V_{a_{i+1}, e}(q)=-Z_{i+1} e^2/\epsilon_0 q^2$ is the Coulomb interaction, and   
$n_B(\omega)=[\exp(\beta \omega)-1]^{-1}$ is the Bose distribution function. 

Correlation effects are described in the low-density limit by the Debye screening, corrections are obtained if  higher order 
Feynman diagrams are taken into account \cite{KKER}. In Debye approximation, the shift of the charged-particle energies is 
$\Delta_i^{\rm corr}=-Z_i \kappa e^2/(8 \pi \epsilon_0)$, and the Coulomb potential is replaced by the statically screened interaction 
$V_{a_{i+1}, e}^{\rm Debye}(r )=-Z_{i+1} e^2/(4 \pi \epsilon_0) \exp(- \kappa r)$,
with the Debye screening parameter $\kappa^2=\sum_i Z_i^2e^2 n_i/(\epsilon_0 k_BT)+\kappa^2_e$,
\begin{equation}
\label{kappa}
 \kappa^2_e=\frac{4 \pi}{k_BT} 2 \left(\frac{2 \pi \hbar^2}{m_e k_BT}\right)^{-3/2}
\frac{e^2}{4 \pi \epsilon_0}
\frac{1}{ \sqrt{\pi}} \int_0^\infty dt \frac{t^{-1/2}}{e^{t-\beta \mu_e}+1}.
\end{equation}
In the region $\Theta \gg 1$, we can solve the BSE (\ref{BSE}) neglecting the effects of degeneracy (Fock shift and 
Pauli blocking). 
The shifts of the bound state, charge $Z_i e$, and of the 
continuum edge of free states because of ionization into the ion with charge $(Z_i+1) e$ and the electron, charge $-e$, 
leads the the IPD in Debye approximation
\begin{equation}
\label{Debye}
\Delta I^{\rm Debye}_i=\kappa (Z_i+1) \frac{e^2}{4 \pi \epsilon_0}.
\end{equation}
Extrapolating this low-density result to higher densities, the Mott point where the ionization potential 
becomes zero (the bound state merges with the continuum of unbound states) is given by 
$\Delta I^{\rm Debye}_i=I_i^{(0)}$.

As example, in this work we discuss carbon ($Z=6$) at high densities/high temperatures where the only bound state to be considered is the ion C$^{5+}$ formed by the nucleus C$^{6+}$ and an electron. The critical 
value for $\kappa$ of the Debye potential where the bound state disappears follows as 
$\kappa_{{\rm C}^{5+}}^{\rm Mott,\,Debye}=5.3075/a_B$. This value corresponds to the Mott density
$n_e^{\rm Mott,\,Debye}=8.318 \times 10^{24}$ cm$^{-3}$ for $T=100$ eV. 
The corresponding carbon mass density is 27.626 g/cm$^{-3}$. 
There, however, we find $\mu^{\rm id}= 79.85$ eV, and the effects of degeneracy become relevant.
(For comparison, for the hydrogen plasma, 
the numerical solution of the BSE for hydrogen atom ($Z_i=0$) gives $\kappa^{\rm Mott, Debye}_{{\rm H}} = 1.159/a_B$.)

The Debye theory cannot be applied to these high densities, in particular the screening 
by the ions is overestimated. The ions are strongly correlated, 
and alternative approaches such as the ion sphere model predict a weaker dependence on density.
Improvements for the contribution of correlations to the IPD, valid also for classical, high density plasmas,
 are given by Stewart and Pyatt \cite{SP66}, see also \cite{Lin17} where the ionic structure factor is included.
We will not discuss correlation effects including higher order Feynman diagrams here any more, 
see \cite{Linnew}, but focus on the effects of degeneracy.

\section{Pauli blocking and perturbation theory}

As before, we take as example a carbon plasma which is the nearly fully ionized so that only the process
C$^{5+} \rightleftharpoons {\rm C}^{6+}+e$ is of relevance.
To investigate the effects of degeneracy, i.e. Pauli blocking and Fock shifts, we simplify the  two-particle equation (\ref{BSE}).
The ions are considered as non-degenerate so that their contribution to the Pauli blocking term is dropped. 
In addition, we replace the dynamical screening by a statically screened (Debye) interaction $V_{{\rm C}^{6+}, e}^{\rm scr}({\bf q}) $ and introduce the quasiparticle shifts $\Delta(k)$ \cite{PRE18},
\begin{eqnarray}
\label{BSE1}
&&\left[E_e(p )+\Delta_e(p )+E_{{\rm C}^{6+}}(k)+\Delta_{{\rm C}^{6+}}(k)
\right] \psi^{5+}_{ n}({\bf p},{\bf k})\nonumber\\&&
+[1-f_e(p )] \sum_{\bf q}V_{{\rm C}^{6+}, e}^{\rm scr}({\bf q}) \psi^{5+}_{ n}({\bf p}+{\bf q},{\bf k}-{\bf q})=
E^{5+}_{n} \psi^{5+}_{ n}({\bf p},{\bf k})\,.
\end{eqnarray}

In adiabatic approximation, the  motion of electrons is separated from the motion of ions. More systematical, we introduce 
Jacobian coordinates, the center-of mass momentum $\bf P$ and the relative momentum ${\bf p}_{\rm rel}$. 
We use a separation ansatz for the wave function 
$ \psi^{5+}_{n}({\bf p},{\bf k}) = \Phi_{ n}({\bf P})\phi_{ n}({\bf p}_{\rm rel})$. 
The center-of-mass motion is given by 
a plane wave. In limit  $m_e \ll M$ where ${\bf p}_{\rm rel}\approx {\bf p}$, we obtain for the relative motion
\begin{eqnarray}
\label{BSE11}
&&\left[E_e(p )+\Delta_e(p )\right] \phi_{ n}({\bf p})
+[1-f_e(p )] \sum_{\bf q}V_{{\rm C}^{6+}, e}^{\rm scr}({\bf q}) \phi_{ n}({\bf p}+{\bf q})=
E^{5+}_{ n,{\rm rel}} \phi_{ n}({\bf p})
\end{eqnarray}
so that $E^{5+}_{ n}=E_{{\rm C}^{6+}}(k)+\Delta_{{\rm C}^{6+}}(k)+E^{5+}_{ n,{\rm rel}}$.

The quasiparticle shift $\Delta_e(p)=\Delta_e^{\rm corr}+\Delta^{\rm Fock}_e(p)$ contains, in addition to the correlation part 
(which can be taken as Debye shift or its improvements),
the Fock shift of an electron with momentum $p$, 
\begin{equation}
\label{HF0}
\Delta^{\rm Fock}_e(p )=\sum_q \frac{e^2}{\epsilon_0 q^2}f_e({\bf p}+{\bf q}).
\end{equation}

For the continuum edge of free (scattering) states, 
the contribution of the Fock shift $\Delta^{\rm Fock}_{e}(0)$ is given by the value at $p=0$. 
Considering the bound states, the ions C$^{5+}$, of the in-medium Schr{\"o}dinger equation (\ref{BSE1}), two contributions arise owing to the electron degeneracy: 
the Fock shift (\ref{HF0}) which modifies the kinetic energy in the Schr{\"o}dinger equation as well as the Pauli blocking term 
in front of the interaction potential. Within a consistent approach, both terms have to be taken into account because they are of the same order of density and partially compensate each other \cite{KKER,RKKKZ78}.

In contrast to the Debye shift which is momentum independent and may be transposed to the chemical potential,
we have to solve the in-medium  Schr{\"o}dinger equation (\ref{BSE1}) which depends on $T$ and $n_e$ via the Fermi distribution function. This will be done in the following sections \ref{sec:sep} and \ref{sec:num}.
Here we shortly recall the perturbation theory \cite{PRE18}.

The unperturbed solution of the hydrogen-like system ${\rm C}^{6+}+e$ is well known. 
We find the ground state solution 
$E^{(Z-1)+}_0=-Z^2 \frac{e^4}{(4 \pi \epsilon_0)^2} \frac{m_e}{2 \hbar^2}=- 13.602 \,Z^2\,\, {\rm eV}$ with $Z=6$ for the case considered here,
 and the corresponding wave function
\begin{equation}
\label{phi0}
 \phi_0( p)= 8 \sqrt{\pi a_Z^3} \frac{1}{(1+a_Z^2 p^2)^2}, \qquad 
\psi_0(r)=\frac{1}{\sqrt{\pi a_Z^3}}e^{-r/a_Z},
\end{equation}
$a_Z=\frac{4 \pi \epsilon_0}{Ze^2} \frac{\hbar^2}{m_e}=a_B/Z$. 

The Fock shift $ \Delta_0^{\rm bound,\,Fock}$ of the bound state energy results in perturbation theory as 
average of the momentum-dependent Fock shift (\ref{HF0})
with the unperturbed wave function (\ref{phi0}),
\begin{equation}
\label{Fb}
 \Delta_0^{\rm bound,\,Fock}=-\sum_{p,q}\phi^2_0( p) \frac{e^2}{\epsilon_0 q^2} f_e({\bf p}+{\bf q})
=-\frac{32}{\pi } \int_0^\infty dp \frac{p^2 a_Z^3}{(1+a_Z^2 p^2)^4}\Delta^{\rm Fock}_e(p )\,.
\end{equation}
Near the Mott condition where the bound state merges with the scattering states, the bound-state wave function
approximates the free one so that the Fock shifts of bound and scattering states have nearly equal values.
Then, they will not contribute to the shift of the binding energy which is the difference between the bound state energy and the continuum edge of free (scattering) states. We will not consider these contributions 
any further in this work, see Ref.~\cite{PRE18} for evaluation. 
      
Within perturbation theory, the Pauli blocking shift is given by
\begin{equation}
\label{Pb}
 \Delta_0^{\rm Pauli,\,perturb}=-\sum_{p,q}\phi_0( p)\,f_e( p)\, V_{{\rm C}^{6+}, e}(q) \,\phi_0({\bf p}+{\bf q})
=\frac{Z e^2}{4 \pi \epsilon_0} \frac{16 a_Z^2}{\pi} \int_0^\infty f_e( p) \frac{p^2 dp}{(1+a_Z^2 p^2)^3}\,.
\end{equation}
Within this work, we analyze this expression and compare it with more rigorous solutions of the in-medium 
Schr\"odinger equation to show the limits of validity of the result (\ref{Pb}) from perturbation theory.

\begin{figure}[h]
  \centerline{\includegraphics[width=350pt,angle=0]{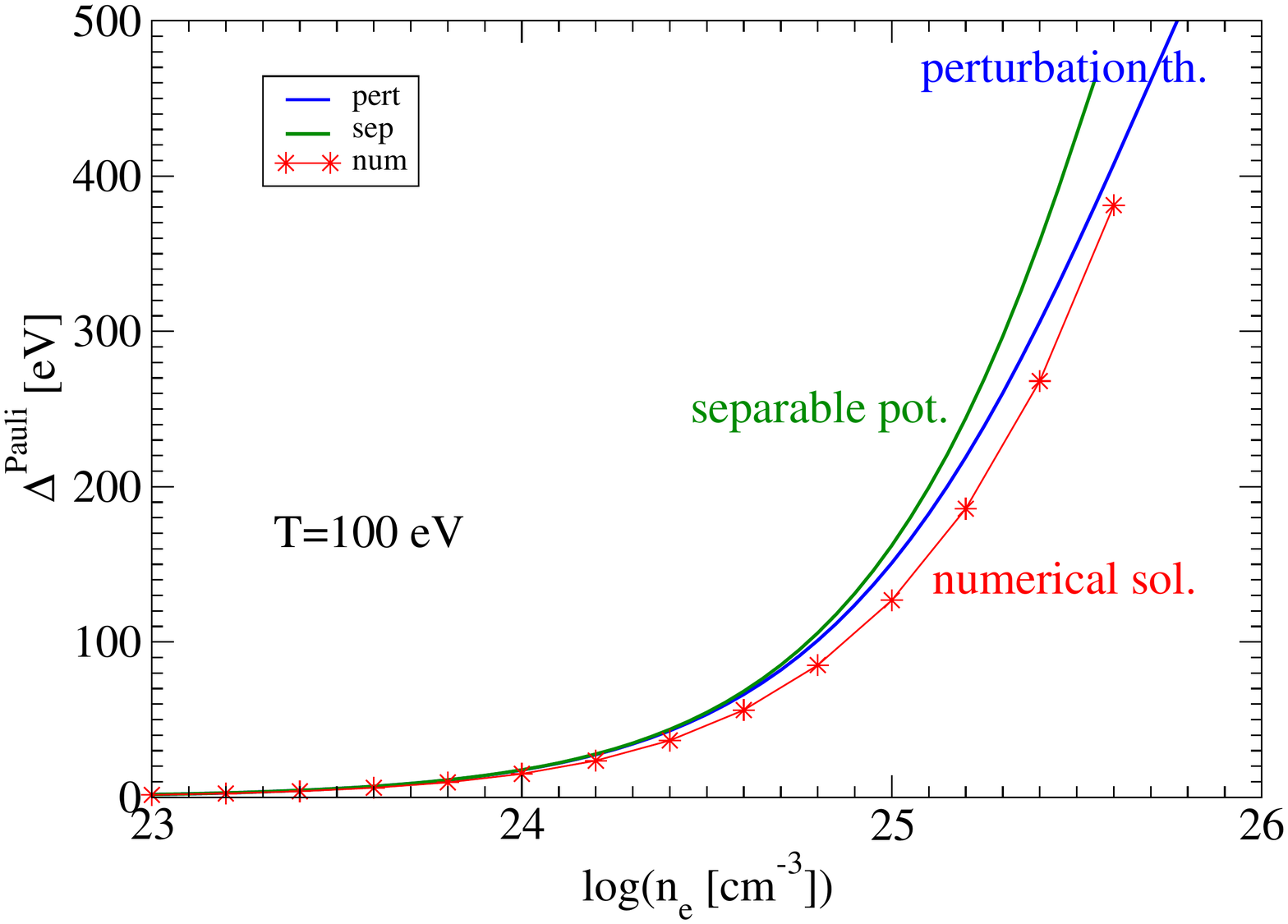}}
  \caption{Pauli blocking shift of the ground state of C$^{6+}$ in the environment of a plasma with the free electron density $n_e$ and given temperature $T=100$ eV.}
\label{fig:1}
\end{figure}

Results for the Pauli blocking shift $\Delta_0^{\rm Pauli,\,perturb}$ are given in Fig. \ref{fig:1} 
for temperature $T= 100$ eV as function of the free electron density $n_e$. For $n_e=10^{25}$ cm$^{-3}$ the value 150.86 eV is obtained.

\section{Separable potentials}
\label{sec:sep}

\subsection{Solution of the in-medium Schr{\"o}dinger equation}

The solution of the Schr\"odinger equation or the related T-matrix becomes simple for separable potentials 
$V^{\rm sep}(p,p') = - \lambda/\Omega\,\, w(p ) w(p')$ (the channel with zero angular momentum is considered).
According to a theorem of Ernst, Shakin and Thaler \cite{EST}, local potentials can be expanded in a series of separable potentials.

We consider the Coulomb interaction (Rydberg units) $V_{{\rm C}^{6+}, e}(q)=-6 e^2/\epsilon_0 q^2= -48 \pi/q^2$ (nuclear charge $Z=6$ for carbon). 
The ground state wave function (\ref{phi0}) and the binding energy 
are reproduced with the separable potential, cf. Ref. \cite{SR87} for the electron-hole system (for nuclear matter this
 Yamaguchi potential is extensively investigated, see
Ref. \cite{R15} and further references given there),
\begin{equation}
\label{Vsep}
 V^{\rm sep}(p,p')=-\frac{16 \pi}{3} \frac{1}{1+p^2/36}  \frac{1}{1+p'^2/36}\,.
\end{equation}

The inclusion of medium effects is simple. The energy eigenvalue follows from the solution of Eq. (\ref{BSE11}).
Considering only the effect of Pauli blocking (the self-energy shift can be added to the kinetic energy $p^2$
 in the denominator), we have to solve
\begin{equation}
 \frac{8}{3 \pi} \int_0^\infty \frac{dp \,p^2}{(1+p^2/36)^2} \frac{1}{p^2-E_0^{\rm sep}}[1-f_e( p)] = 1\,.
\end{equation}
Because the blocking term with $ f_e( p)$ is dependent on $T, \mu_e$, the solution $E_0^{\rm sep}$ also depends on $T, n_e$. 
The result for $T=100$ eV is shown in Fig. \ref{fig:1}. It is not very different from the 
 result of  perturbation theory.

The in-medium ground-state wave function is also immediately obtained,
\begin{equation}
\psi_0^{\rm sep}(p ) \propto \sqrt{1-f_e( p)} \frac{1}{1+p^2/36} \frac{1}{p^2-E_0^{\rm sep}}.
\end{equation}
The Fourier transformation gives the  wave function in position space representation,
\begin{equation}
\label{psir}
\psi_0^{\rm sep}(r ) \propto \frac{1}{r} \int_0^\infty dp\,p \sin(rp)\,\psi_0^{\rm sep}(p ).
\end{equation}
All expressions are simplified for $T=0$ where the Fermi distribution function becomes the step function which jumps from 1 to 0 at the Fermi momentum $p_F=(3 \pi^2 n_e)^{2/3}$.


%

%

\begin{figure}[h]
  \centerline{\includegraphics[width=350pt,angle=0]{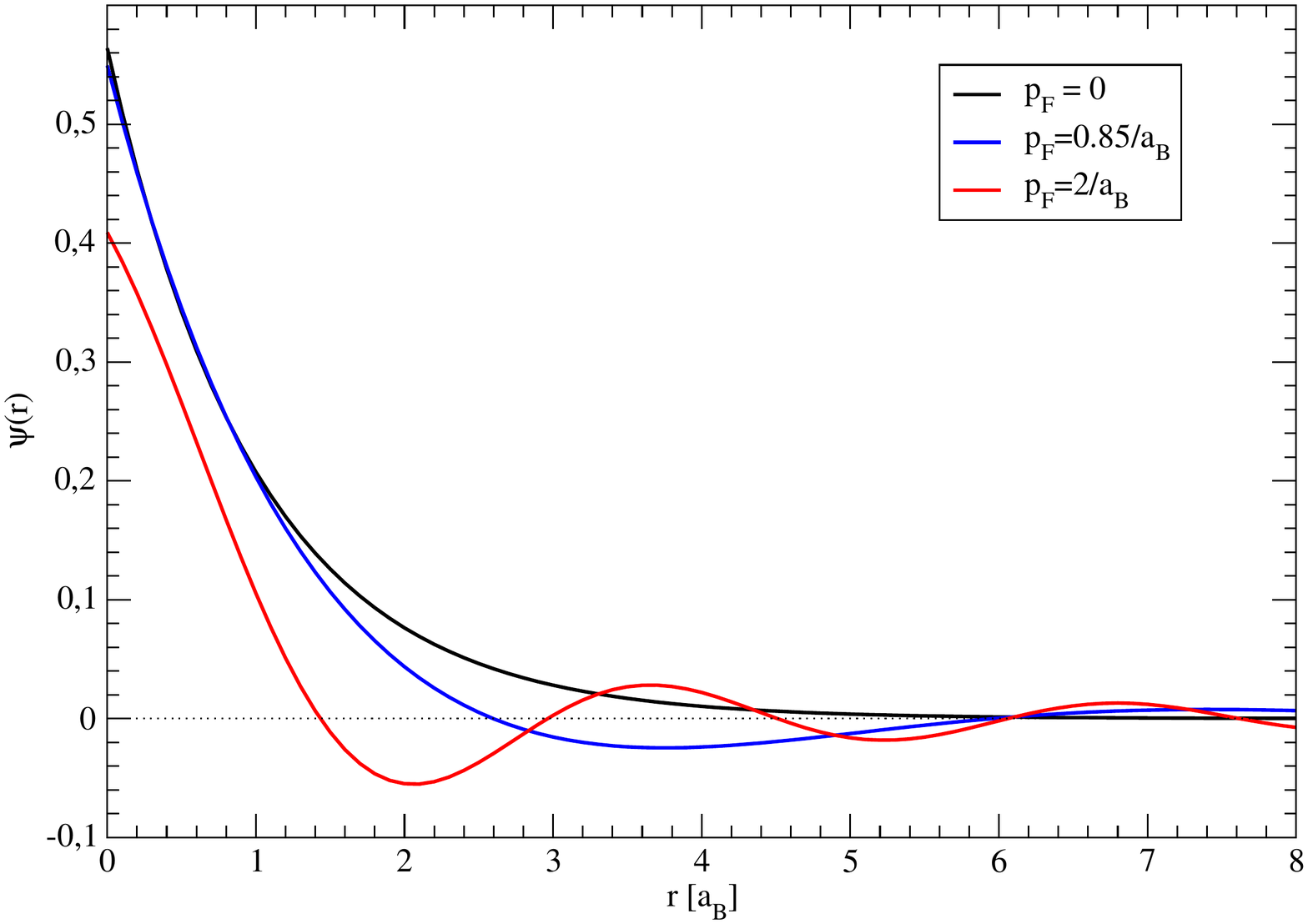}}
  \caption{Normalized bound state wave function $\psi_0^{\rm sep}( r)$ (\ref{psir}) at $T=0$ for different free electron densities as given by the corresponding Fermi momenta. For the interaction between the C$^{6+}$ ion and the electron, 
  the separable potential (\ref{Vsep}) has been used.}
\label{fig:2}
\end{figure}

The medium modification of the ground-state  normalized wave function is shown in Fig. \ref{fig:2} at $T=0$ as a function of the free electron density. With increasing density, the well-known wave function for the localized hydrogen-like ground state becomes oscillating (Friedel oscillations) approaching the 
free $s$ state at the Fermi energy in the high-density limit.

\begin{figure}[h]
  \centerline{\includegraphics[width=350pt,angle=0]{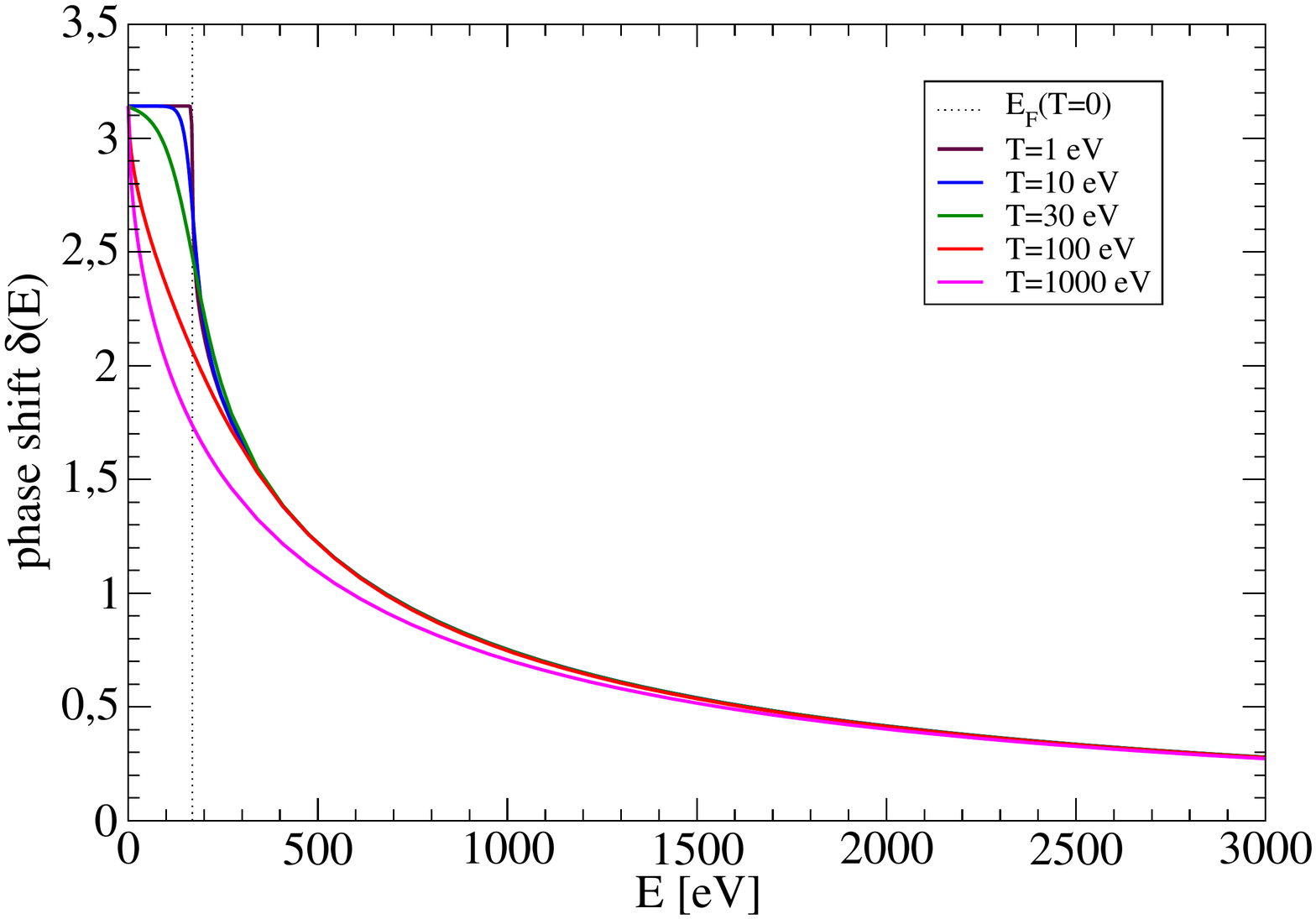}}
  \caption{Phase shift $\delta(E)$ as function of $E$ for different temperatures $T=1,\,10,\,30,\,100,\,1000$ eV. 
Free electron density $n_e = 10^{25}$ cm$^{-3}$. Fermi energy $E_F=169.25$ eV.}
\label{fig:3}
\end{figure}

An important feature of the separable potential is that also the scattering states are immediately obtained. 
This allows to implement the continuum correlations. 
For the potential
\begin{equation}
\label{vsep}
 V^{\rm sep}(p,p')=-\frac{32 \pi a_Z}{\Omega} \,\frac{\hbar^2}{2 m_e}\,  \frac{1}{1+a_Z^2 p^2}\,\frac{1}{1+a_Z^2 {p'}^2} \,.
\end{equation}
we have 
\begin{eqnarray}
\label{I(E)}
 I(E)&=&\frac{16 a_Z}{\pi} \frac{2 m_e}{\hbar^2}\int_0^\infty dp \frac{p^2}{(1+a_Z^2 p^2)^2}\frac{1}{(2 m_e/\hbar^2)E-p^2}
[1-\hat f_e(E_p)]\nonumber\\&&
=\frac{16 a_Z}{\pi} \frac{2 m_e}{\hbar^2} \frac{\pi}{4  a_Z (1+ {\rm i} a_Z[(2 m_e/\hbar^2)E]^{1/2})^2}[1-\hat f_e(E)]\nonumber\\&&
+\frac{16 a_Z}{\pi} \frac{2 m_e}{\hbar^2}\int_0^\infty dp \frac{p^2}{(1+a_Z^2 p^2)^2}\frac{1}{(2 m_e/\hbar^2)E-p^2}
[\hat f_e(E)-\hat f_e(E_p)]
\end{eqnarray}
The scattering phase shifts follow from 
\begin{equation}
 \delta(E)=\arctan\left(\frac{- {\rm Im}\,I(E) }{1-{\rm Re}\,I(E)}\right)\,.
\end{equation}
As example, they are shown in Fig. \ref{fig:3} for the free electron density $n_e = 10^{25}$ cm$^{-3}$ and for the temperatures 
$T=1,\,10,\,30,\,100,\,1000$ eV (the corresponding chemical potentials are respectively $\mu^{\rm id}_e=169.25,\, 168.77,\,164.62,\,107.74,\,-2930.72     $ eV). 
Whereas in the high-temperature limit the phase shifts are decreasing with increasing energy $E$, the phase shift at zero temperature remains constant, $\delta(E)=\pi$, for energies below the Fermi energy. 
This is a consequence of the Pauli blocking. We conclude that in general the contributions of the scattering states 
to the spectral function and the partial densities cannot be neglected. 
A similar behavior was also obtained for nuclear matter \cite{Phaseshifts}.

Note that we can also consider finite Debye screening which modifies the separable potential 
(because it is constructed to reproduce the bound-state wave function), 
but the general qualitative behavior of the 
wave function in position space as well as in momentum space is not changed if screening is taken into account.\\

\subsection{Density of states}
\label{dos}

The density of states is related to the spectral function $A_e(1, \omega)$ and the self-energy $\Sigma_e(1,z)$ according to
\begin{equation}
\label{dosa}
 D_e(\omega)=\frac{1}{2 \pi \Omega}\sum_1 A_e(1, \omega)=\frac{1}{\pi \Omega}\sum_1 
\frac{{\rm Im}\,\Sigma_e(1,\omega+i0)}{[\omega-E_e(1)- {\rm Re}\,\Sigma_e(1,\omega)]^2+[{\rm Im}\,\Sigma_e(1,\omega+i0)]^2}\,.
\end{equation}
The self-energy follows from the T-matrix as
\begin{equation}
\label{seT}
 \Sigma_e(1,i z_\nu)=\frac{1}{\beta}\sum_{2,\Omega_\mu} T(1,2;1,2;i\Omega_\mu)\frac{1}{i \Omega_\mu-iz_\nu-E_i(2)}
\end{equation}
where $\Omega_\mu\,\,(z_\nu)$ are bosonic (fermionic) Matsubara frequencies, 2 denotes the quantum state of the ion 
(momentum $k$ and spin). We consider here only the contribution owing to the electron-ion interaction, 
the electron-electron interaction gives no bound state and is well investigated for plasmas elsewhere \cite{KKER,Dornheim}.

The electron-ion interaction can be treated in different approximations, neglecting ion-ion correlations or 
taking into account ion-ion correlations as ionic structure factor, and accounting for the motion of the ions 
or treating them in adiabatic approximation.

As before, we consider a separable potential, Eq. (\ref{Vsep}), 
where the solution of the Schr\"odinger equation can be given explicitly. 
Considering independent interactions with the ions, we have after introducing total ($\bf P=p+k$) and relative momenta
(${\bf p}_{\rm rel}=(M_i {\bf p}-m_e {\bf k})/(M_i+m_e)$) for the T-matrix (spin variables define the channel $\gamma$)
\begin{equation}
 T({\bf p},{\bf k};{\bf p}'{\bf k}',z)=
\frac{\lambda}{\Omega}w(p_{\rm rel} ) w(p_{\rm rel}')\delta_{\bf P, P'}
\frac{1}{1+\frac{\lambda}{\Omega} \sum_{p''}w^2(p_{\rm rel}'')\frac{1-f_e(p_{\rm rel}'')}{z-E_e(p_{\rm rel}'')-E_i(P)}}
\end{equation}
(we used ${\bf P}''={\bf P}$ and the adiabatic approximation ${\bf p}''= {\bf p}_{\rm rel}'',\,\,{\bf k}''={\bf P}$ in the last term); ion-ion correlations are neglected.
The spectral representation of the T-matrix reads
\begin{equation}
 T({\bf p},{\bf k};{\bf p}'{\bf k}',i\Omega_\mu)=
\frac{\lambda}{\Omega}w(p_{\rm rel} ) w(p_{\rm rel}')\delta_{\bf P, P'}+\int_{-\infty}^\infty \frac{d \omega}{\pi}
\frac{1}{i \Omega_\mu-\omega} {\rm Im}\, T({\bf p},{\bf k};{\bf p}'{\bf k}',\omega-i0).
\end{equation}
The summation over $\Omega_\mu$ in Eq. (\ref{seT}) can be performed with 
\begin{equation}
 \frac{1}{\beta}\sum_{\Omega_\mu} \frac{1}{i \Omega_\mu-\omega} \frac{1}{i \Omega_\mu-iz_\nu-E_i(2)}
=\frac{1}{iz_\nu+E_i(2)-\omega}\left[\hat f_i(E_i(2))+\hat g_{ei}(\omega)\right].
\end{equation}
In the low-density limit, the Bose distribution $\hat g_{ei}(\omega)$ containing $\mu_i+\mu_e$ can be neglected.
Neglecting Pauli blocking in the low-density limit, with the C$^{6+}-e$ interaction we have (Rydberg units)
\begin{equation}
 {\rm Im} \Sigma_e(p,\omega+i0)=\frac{16 \pi}{3} \frac{1}{(1+p^2/36)^2} \sum_2 \hat f_i(E_i(2)) \,\,
{\rm Im} \frac{1}{1-4/(1+(-\omega-i0+E_i(2))^{1/2}/6)^2}
\end{equation}
where
\begin{equation}
\label{Im}
{\rm Im} \frac{1}{1-\frac{4}{(1+\sqrt{-\omega-i0+E_i(2)}/6)^2}}=\left\{\begin{array}{ll}
        \pi \delta[1-4/(1+\sqrt{-\omega+E_i(2)}/6)^2] &  \,\,{\rm if}\, \omega-E_i(2)<0,\\
\frac{1728 \sqrt{\omega-E_i(2)}}{11664-72 (\omega-E_i(2))+(\omega-E_i(2))^2} & \,\, {\rm if}\,\omega-E_i(2)>0
       \end{array}
\right\}.
\end{equation}

%
%
%
%
%
%

Now we can evaluate the DOS using (\ref{dosa}). The ${\rm Re} \Sigma_e(p,\omega)$ is obtained from the Kramers-Kronig relation, 
taking the Hartree-Fock shift into account. 
If we use the expansion for small ${\rm Im} \Sigma_e(p,\omega+i0)$, see, e.g., Refs. \cite{PRE18,R15,SRS}, we have a first 
contribution originated from the single quasiparticle contribution ("free" electron density). The correlated part
\begin{equation}
 D_e^{\rm corr} (\omega) = \frac{1}{\pi \Omega}\sum_1 
{\rm Im}\,\Sigma_e(1,\omega+i0)\frac{d}{d \omega}\frac{{\cal P}}{\omega-E_e(1)}
\end{equation}
with the principal value ${\cal P}$ needs some care treating the behavior at $\omega=E_e(1)$ (the singularity is regularized). The bound state contribution
is not influenced by the singularity, with $\delta[1-4/(1+(-\omega+E_i(2))^{1/2}/6)^2]=\delta(\omega-E_i(2)+36) 12 \sqrt{2}/(36+
E_e(p)+E_i(2))^2$ the integrals over $\bf p,k$ can be performed.
The result is 
\begin{equation}
 D_e^{\rm bound} (\omega) = \frac{1}{ \Omega}\sum_{{\bf P},{\bf p}_{\rm rel}} |\psi_0(p_{\rm rel}|^2 
f_i[E_i({\bf P}-{\bf p}_{\rm rel})]  
\delta\left[\omega + E_i({\bf P}-{\bf p}_{\rm rel})+36-\hbar^2  P^2/(2 M)\right]
\end{equation}
(with the carbon ion mass $M\approx 12 000$ going to Rydberg units). We find with $\theta(x) = 1, x>0$, $\theta(x) = 0, x<0$:
\begin{eqnarray}
 D_e^{\rm bound} (\omega) &=& \frac{M}{ 27 \pi^3}\int P dP \int  d p_{\rm rel} \frac{p_{\rm rel}}{(1+p_{\rm rel}^2/36)^4}
e^{(36+\omega-P^2/(2M)+\mu_i)/T} \nonumber \\&& \times\left(1-\theta \left[\left(\frac{M}{Pp_{\rm rel}}(\omega+36) 
-\frac{P}{4 M p_{\rm rel}}+\frac{p_{\rm rel}}{2 P}\right)^2-1\right]\right)
\end{eqnarray}
which gives a broadened peak near $\omega = -36 =-490.5$ eV with the width of about 1 eV at $T=100$ eV, $n_e=10^{25}$ cm$^{-3}$.
A better approach should not only consider the inelastic collision of the electron with the ions, 
but it has take also into account the ionic structure factor.

For the contribution of scattering states we have to consider the singularity at $\omega-E_e(1)=0$. We have
\begin{equation}
 D_e^{\rm scat} (\omega) = \frac{8 n_i}{ 3 \pi} 1728 \int \frac{p^2 dp}{(1+p^2/36)^2} \frac{1}{(\omega-p^2)^2} d_e(\omega), 
\qquad  d_e(\omega)=\frac{\sqrt{\omega}}{11664-72 \omega+\omega^2}.
\end{equation}
We consider the following decomposition to regularize the integral: In the integral expression, we replace 
$d_e(\omega)$ by $d_e(\omega)-d_e(p^2)-d_e'(\omega) (\omega-p^2)$. This integral is regular and can be performed.
We have a second integral where we replace $d_e(\omega)$ by $d_e'(\omega) (\omega-p^2)$. Now we can perform the principal 
value integral over $p$ with the result ${\rm Re}[ 54 \pi/(6+i \sqrt{\omega})^2]= 54 \pi(36-\omega)/[(36-\omega)^2+144 \omega]$.
The last contribution, where $d_e(\omega)$ is replaced by $d_e(p^2)$, is singular, but is already compensated by the 
self-energy shift of the quasiparticle energy. Now the contribution $D_e^{\rm scat} (\omega)$ to the DOS can be evaluated.
It gives an enhancement for $\omega>0$ near the continuum edge.
To consider the effects of Pauli blocking at increasing densities, we have to analyze $I(E)$ defined in Eq. (\ref{I(E)}).\\

\subsection{Many-ion system}

The independent interaction with single ions is not sufficient to describe WDM. Instead, we have to consider multiple scattering at all ions, at positions ${\bf R}_i$ (in adiabatic approximation the ion configuration is fixed).
To introduce the ion distribution, we perform the translation of the non-local potential $V^{\rm sep}({\bf r},{\bf r}') = \sum_{p,p'} \exp[i({\bf p}{\bf r}-{\bf p}'{\bf r}')] V^{\rm sep}( p,p')$ so that $V^{\rm ions}({\bf r},{\bf r}';\{{\bf R}_i \})
=\sum_i V^{\rm sep}({\bf r}-{\bf R}_i,{\bf r}'-{\bf R}_i)$, or
\begin{equation}
 V^{\rm ions}({\bf p},{\bf p}';\{{\bf R}_i \})=\sum_i^{N_i} e^{i({\bf p}-{\bf p}'){\bf R}_i} V^{\rm sep}( p,p')
=S_{ii}({\bf p}-{\bf p}') V^{\rm sep}( p,p')
\end{equation}
with the ionic structure factor $S_{ii}({\bf q})$. The Schr{\"o}dinger equation (Rydberg units) reads
\begin{equation}
 (p^2-E) \psi_\nu({\bf p})-\frac{\lambda}{\Omega}\sum_i e^{i{\bf p}\cdot {\bf R}_i} w(p) 
\sum_{p'}e^{-i{\bf p}'\cdot {\bf R}_i} w(p')  \psi_\nu({\bf p}')
=(p^2-E) \psi_\nu({\bf p})-\frac{\lambda}{\Omega}\sum_i e^{i{\bf p}\cdot {\bf R}_i} w(p) c_{i,\nu} =0
\end{equation}
so that for $\psi_\nu({\bf p})=\sum_i c_{i,\nu} w(p)$
\begin{equation}
 c_{i,\nu}=\sum_{p,j}e^{-i{\bf p}\cdot ({\bf R}_i-{\bf R}_j)} \frac{\lambda}{\Omega} \frac{w^2(p)}{p^2-E_\nu}c_{j,\nu}
=\sum_{j}a_{i,j}(E_\nu) c_{j,\nu}\,.
\end{equation}
More general, $\lambda$ and $w(p)$ may also depend on the ion position $i$. For the separable potential (\ref{vsep}) we find
\begin{equation}
 a_{i,j}(E)=\frac{\lambda}{2 \pi^2 R_{ij}}\int_0^\infty \frac{p\, dp}{(1+a_Z^2 p^2)^2(p^2-E)} \sin(p R_{ij})\,
\end{equation}
where for the C$^{6+}$ ion $\lambda=16 \pi/3$ and $a_Z=1/6$  holds.
We find
\begin{eqnarray}
 &&\int_0^\infty \frac{p\, dp}{(1+a_Z^2 p^2)^2(p^2-E)} \sin(p R_{ij})\nonumber \\&&
=-\frac{\pi R_{ij}}{4 a_Z (1+a_Z^2 E)} e^{-R_{ij}/a_Z}-\frac{\pi}{2 (1+a_Z^2 E)^2}\left(e^{-R_{ij}/a_Z}-
e^{-\sqrt{-E}R_{ij}}\right)\,.
\end{eqnarray}

With these matrix elements, we solve the equation for $c_{i,\nu}$. The $N_i$ energy eigenvalues $E_\nu$ follow from 
$|\delta_{ij}-a_{ij}(E_\nu)|=0$, and the corresponding eigenvectors $c_{i,\nu}$ determine the corresponding wave function.
For periodic ion configurations, the band structure is obtained.
In particular, for a sc lattice we have the solution $c_{i,\nu}\propto e^{i {\bf k}_\nu \cdot {\bf R}_i}$ 
with ${\bf k}_\nu$ being a vector of the reciprocal lattice. The bandwidth follows from the solution of
\begin{equation}
\sum_{\langle i,j \rangle} e^{i {\bf k}_\nu \cdot ({\bf R}_j-{\bf R}_i)} a_{i,j}(E_\nu)+1-a_{i,i}(E_\nu)=0.
\end{equation}
Calculations for C at 50 g/cm$^{3}$ (lattice parameter $1.39 \,a_B$) gives the bandwidth $W=26.53$ eV, 
and for 20 g/cm$^{3}$  (lattice parameter $1.887\, a_B$) we find $W=1.79$ eV.
For disordered systems with the distribution of ${\bf R}_i$ according to the structure factor $S_{ii}({\bf q})$, 
one has to solve a matrix equation.

\section{Numerical solution of the Pauli blocking shift}
\label{sec:num}

We come back the the screened Coulomb potential with the Debye screening parameter $\kappa$, Eq. (\ref{kappa}). We present numerical solutions of the hermitean form of the wave equation (\ref{BSE11})
\begin{eqnarray}
\label{BSE1n}
&&\frac{\hbar^2}{2m_e} p^2  \psi_0^{\rm herm}({\bf p})
-\sqrt{1-f_e(p )} \,\,\sum_{\bf q} \frac{Ze^2}{\epsilon_0 \Omega ({\bf p}- {\bf q})^2+\kappa^2}\sqrt{1-f_e(q )} \,\, \psi_0^{\rm herm}({\bf q})=
E \psi_0^{\rm herm}({\bf p})\,.
\end{eqnarray}
With a small value of $\kappa$ we avoid the singularity at $\bf p=q$ 
(the infinite-range Coulomb potential, where phase shifts cannot be introduced in the standard way,  
is replace by a finite-range Debye potential).
As example, we choose $\kappa=0.5/a_B$ as a formal parameter (it corresponds to $n_e= 7.08 \times 10^{22}$ cm$^{-3}$ at $T=100$ eV). The solution of the Schr\"odinger equation for $Z=6$ without Pauli blocking gives the ground state energy $E_{0,0}=-412.949$ eV.

To find the ground state energy and wave function for arbitrary $T,n_e$ taking Pauli blocking into account, 
an eigenvalue problem was solved after 
discretization of Eq. (\ref{BSE1n}) in momentum space. 
The effect of Pauli blocking leads to the energy shift shown in Fig. \ref{fig:1} for $T=100$ eV as function of 
the free electron density $n_e$. It is a little bit smaller compared to the perturbation theory because the 
deformation of the wave function is taken into account, what leads to a lower energy eigenvalue as known from variational calculations. 
In addition, numerical solutions for the scattering phase shifts and the related quantities as discussed above 
for the separable potential may be obtained for the Debye potential, but this goes beyond the scope of the present work.

\section{Conclusions and outlook}

Our main issue was to improve the PIP model taking scattering states into account, what is performed by decomposition of the 
density into an uncorrelated part and a correlated part, i.e. a cluster decomposition of the density as function of 
 the chemical potentials and $T$. In addition, we calculated the effect of Pauli blocking which is essential at high densities where the electrons become degenerate. A perturbative treatment was compared to a numerical solution and an analytical approach 
based on the separable potential approach. The latter can be used to find explicit solutions for the density of states, 
including the contribution of scattering states.

We focus in this work on the effect of Pauli blocking. Additional effects described by the in-medium Schr{\"o}dinger equation
(\ref{BSE}), (\ref{BSE1}) are discussed in Ref. \cite{PRE18}. With the present contribution, we intend to derive 
results for the Pauli blocking which go beyond perturbation theory as used in \cite{PRE18}.

Controversies such as the treatment of the  $K$-edge shifting in strongly degenerate 
 systems \cite{Hu,Rosmej}, where $\mu_e \approx E_F$, may be resolved within the quantum statistical approach. 
The optical response function should be calculated taking into account the contribution of  scattering phase shifts, 
see Fig. \ref{fig:3} and Ref. \cite{Phaseshifts}. 
For the strongly degenerate electron gas, bound-state like contributions do not disappear  
 if the bound state merges with the continuum of scattering states. At zero temperature, bound-state like correlations in the continuum give a significant contribution to the correlated density, see Fig. \ref{fig:3}.

For the Pauli blocking, the phase space occupation in Eqs. (\ref{BSE}), (\ref{BSE1}) was taken according to the ideal Fermi gas.
Within a self-consistent approach, the phase space occupation should be given by the spectral function to be calculated. 
The modification of 
the single-particle occupation numbers in phase space because of correlations has been considered in Ref. \cite{R15}. 

An alternative approach to WDM is given by the density-functional theory \cite{Vinko12,VCW14,Vinko15,Vinko18} 
where the single-particle density of states is evaluated. In Ref.  \cite{PNP16}, orbital-free 
molecular dynamics is performed, and the sensitivity of the equations of state, obtained there, to the choice 
of exchange-correlation functionals is investigated.
There is no simple relation between DFT calculations and the PIP model, because the DFT approach considers 
electrons moving in a mean field like uncorrelated quasiparticles. Correlations such as multiple occupation 
of a given ion are not 
included in these considerations. In contrast, the Hubbard model which describes electron-electron correlations 
beyond a mean-field approach, can describe the different occupation numbers of ions. 
It is problematic to perform DFT calculations in the low-density region where the plasma is described adequately
by the PIP model and the Saha equations. DFT calculations may become relevant in the high-density region where 
all electrons are nearly free so that a mean-field approach with appropriate approximations 
for exchange and correlation energy density is justified. Work in this direction is in progress \cite{Mandy}.

\section{ACKNOWLEDGMENTS}
The author would like to express thanks to M. Bethkenhagen, Ch. Lin, W.-D. Kraeft, R. Redmer, and H. Reinholz (Rostock), 
as well as D. Blaschke (Wroclaw) and T. Doeppner (Livermore) for stimulating discussions.


\begin{thebibliography}{10}
\footnotesize


\bibitem{WarmDM}
 R. W. Lee, D. Kalantar, and J. Molitoris, {\it Warm Dense Matter:
An Overview}, in Livemore, UCRL-TR-203844, https://e-reports-ext.llnl.gov/pdf/307164.pdf (2004)

   \bibitem{KKER}
W.-D. Kraeft, D. Kremp, W. Ebeling and G. R\"opke, {\it Quantum Statistics of Charged Particle Systems} 
(Akademie-Verlag Berlin, 1986).

  \bibitem{KSK}
D. Kremp, M. Schlanges, and W.-D. Kraeft, {\it Quantum Statistics of Nonideal Plasmas} 
(Springer-Verlag Berlin Heidelberg, 2005).

\bibitem{EST}
D. J. Ernst, C. M. Shakin, and R. M. Thaler, Phys. Rev. C {\bf 8}, 46 (1973).

\bibitem{SR87}
M. Schmidt and G. R\"opke, Phys. Stat. Sol. (b) {\bf 139}, 441 (1987);\\
M. Schmidt, T. Janke,  and R. Redmer, Contrib. Plasma  Phys. {\bf 29}, 431 (1989). 

\bibitem{R15}
G. R\"opke, Phys. Rev. C {\bf 92}, 054001 (2015).
 
\bibitem{Militzer}
K. P. Driver, F. Soubiran, and B. Militzer,
 Phys. Rev. E {\bf 97}, 063207 (2018);\\
K. P. Driver and B. Militzer,
 \pr{Lett.}{108}{115502}{2012};\\
S. Zhang, K. P. Driver, F. Soubiran, and B. Militzer,
Phys. Rev. E {\bf 96}, 013204 (2017).

\bibitem{Dornheim}
T. Dornheim, S. Groth, and M. Bonitz, Phys. Rep. {\bf 744}, 1 (2018).

\bibitem{PRE18}
G. R\"opke \etal, submitted to Phys. Rev. E, [arxiv:1811.12912].

 \bibitem{Hoarty13}
 D. J. Hoarty \etal, 
 \pr{Lett.}{110}{265003}{2013}.
 
  \bibitem{ciricosta12}
 O. Ciricosta \etal, 
 \pr{Lett.}{109}{065002}{2012}.
 
   \bibitem{ciricosta16}
 O. Ciricosta \etal, 
Nat. Commun. {\bf 7}, 11713 (2016).

  \bibitem{Vinko18}
M. F. Kasim, J. S. Wark, and S. M. Vinko,
Scient. Rep. {\bf 8}, 6276 (2018).

 \bibitem{Fletcher14}
 L. B. Fletcher \etal, 
\pr{Lett.}{112}{145004}{2014}. 
 
\bibitem{Kraus16}
D. Kraus \etal, 
Phys. Rev. E {\bf 94}, 011202(R) (2016).

\bibitem{CFT15}
 A. Calisti, S. Ferri, and B. Talin, 
J. Phys. B: At. Mol. Opt. Phys. {\bf 48}, 224003 (2015). 

\bibitem{Crowley14}
 B. J. B. Crowley, 
\hedp{13}{84}{2014}.

\bibitem{Lin17}
C. Lin, G. R\"opke, W. D. Kraeft, and H.
Reinholz, Phys. Rev. E {\bf 96}, 013202 (2017).

\bibitem{Udo}
G. R\"opke,  {\it Correlation and Clustering in Dilute Matter}, in: W.U. Schr\"oder (ed.),
{\it Nuclear Particle Correlations and Cluster Physics} 
(World Scientific, 2017) [arXiv:1703.06734]. 


 \bibitem{RKKKZ78}
 G. R\"opke, K. Kilimann, D. Kremp, W.D. Kraeft, and R. Zimmermann, 
 Phys. Stat. Sol. (b) {\bf 88}, K59 (1978);\\
 R. Zimmermann, K. Kilimann, W.D. Kraeft, D. Kremp, and G. R\"opke, 
 Phys. Stat. Sol. (b) {\bf 90}, 175 (1978).


  \bibitem{SP66}
 J. C. Stewart and K. D. Pyatt, Jr., 
 \astropj{144}{1203}{1966}.

\bibitem{Linnew}
C. Lin {\it et al.}, in preparation.

\bibitem{Phaseshifts}
G. R\"opke, J. Phys.: Conf. Series {\bf 569}, 012031 (2014).

\bibitem{SRS}
M. Schmidt, G. R\"opke, and H. Schulz, Ann. Phys. {\bf 202}, 57 (1990).

 
 \bibitem{Hu}
S. X. Hu, Phys. Rev. Lett. {\bf 119}, 065001 (2017);\\
C. A. Iglesias and P. A. Sterne,  Phys. Rev. Lett. {\bf 120}, 119501 (2018);\\
S. X. Hu,  Phys. Rev. Lett. {\bf 120}, 119502 (2018).

  
\bibitem{Rosmej}
 F. B. Rosmej, 
J. Phys. B: At. Mol. Opt. Phys.
{\bf 51} 09LT01 (2018).


  \bibitem{Vinko12}
 S. M. Vinko \etal, 
 Nature {\bf 482}, 59 (2012).

   \bibitem{VCW14}
 S. M. Vinko, O. Ciricosta, and J. S. Wark, 
 Nat. Commun. {\bf 5}, 3533 (2014).
 
  \bibitem{Vinko15}
 S. M. Vinko \etal, 
 Nat. Commun. {\bf 6}, 6397 (2015).
 

\bibitem{PNP16}
J.-F. Danel, L. Kazandjian, and R. Piron, Phys. Rev. E {\bf 98}, 043204 (2018).

\bibitem{Mandy}
M. Bethkenhagen,  private communication.


\end{thebibliography}
\end{document}